\documentclass{jfm}

\usepackage{graphicx}
\usepackage{natbib}
\usepackage{color}

\ifCUPmtlplainloaded \else
  \checkfont{eurm10}
  \iffontfound
    \IfFileExists{upmath.sty}
      {\typeout{^^JFound AMS Euler Roman fonts on the system,
                   using the 'upmath' package.^^J}%
       \usepackage{upmath}}
      {\typeout{^^JFound AMS Euler Roman fonts on the system, but you
                   dont seem to have the}%
       \typeout{'upmath' package installed. JFM.cls can take advantage
                 of these fonts,^^Jif you use 'upmath' package.^^J}%
      }
  \else
  \fi
\fi

\ifCUPmtlplainloaded \else
  \checkfont{msam10}
  \iffontfound
    \IfFileExists{amssymb.sty}
      {\typeout{^^JFound AMS Symbol fonts on the system, using the
                'amssymb' package.^^J}%
       \usepackage{amssymb}%

      }{}
  \fi
\fi

\ifCUPmtlplainloaded \else
  \IfFileExists{amsbsy.sty}
    {\typeout{^^JFound the 'amsbsy' package on the system, using it.^^J}%
     \usepackage{amsbsy}}
    {}
\fi

\newsavebox{\astrutbox}
\sbox{\astrutbox}{\rule[-5pt]{0pt}{20pt}}

\newcommand\thalf{\ensuremath{{\textstyle\frac{1}{2}}}}

\title[Highly focused supersonic microjets]{Highly focused supersonic microjets:\\numerical simulations}

\author[I.R. Peters, Y. Tagawa, et al.]
{I\ls V\ls O\ns R.\ns P\ls E\ls T\ls E\ls R\ls S$^1$,\ns
Y\ls O\ls S\ls H\ls I\ls Y\ls U\ls K\ls I\ns T\ls A\ls G\ls A\ls W\ls A$^1$,\break
N\ls I\ls K\ls O\ls L\ls A\ls I\ns O\ls U\ls D\ls A\ls L\ls O\ls V$^1$,\ns
C\ls H\ls A\ls O\ns S\ls U\ls N$^1$,\break
A\ls N\ls D\ls R\ls E\ls A\ns P\ls R\ls O\ls S\ls P\ls E\ls R\ls E\ls T\ls T\ls I$^{1,2}$,\ns
D\ls E\ls T\ls L\ls E\ls F\ns  L\ls O\ls H\ls S\ls E$^1$\break
\and
D\ls E\ls V\ls A\ls R\ls A\ls J\ns v\ls a\ls n\ns d\ls e\ls r\ns M\ls E\ls E\ls R$^1$}

\affiliation{$^1$Department of Applied Physics and J.M. Burgerscentre for Fluid Dynamics, \break University of Twente,
PO Box 217, 7500 AE Enschede, The Netherlands \break
$^2$Department of Mechanical Engineering, John Hopkins University, Baltimore, MD 21218, USA}

\pubyear{????}
\volume{???}
\pagerange{???--???}
\date{?? and in revised form ??}

\begin{document}

\maketitle

\begin{abstract}
By focusing a laser pulse inside a capillary partially filled with liquid, a vapour bubble is created which emits a pressure wave. This pressure wave travels through the liquid and creates a fast, focused axisymmetric microjet when it is reflected at the meniscus. We numerically investigate the formation of this microjet using axisymmetric boundary-integral simulations, where we model the pressure wave as a pressure pulse applied on the bubble. We find a good agreement between the simulations and experimental results in terms of the time evolution of the jet and on all parameters that can be compared directly. We present a simple analytical model that accurately predicts the velocity of the jet after the pressure pulse and its maximum velocity.
\end{abstract}


\section{Introduction}
In recent experiments by \cite{Tagawa2012}, it was found that  microscopic jets that travel at a speeds up to $850~\mathrm{m/s}$ can be created by focusing a laser pulse inside a liquid-filled capillary that is open at one end. Besides the high velocity, the jets were found to be highly reproducible and controllable. The laser pulse used in the experiments, which has an energy of the order of $100~\mathrm{\mu J}$, results in the formation of a vapour bubble accompanied by a pressure wave \citep{Bell1967,Felix1971}. This pressure wave is reflected at the free surface, where the jet is formed. The shape of the free surface was found to play a crucial role in the formation of the jet, as it is responsible for focusing the liquid into a jet.

\begin{figure}
  \centerline{\includegraphics[width=10cm]{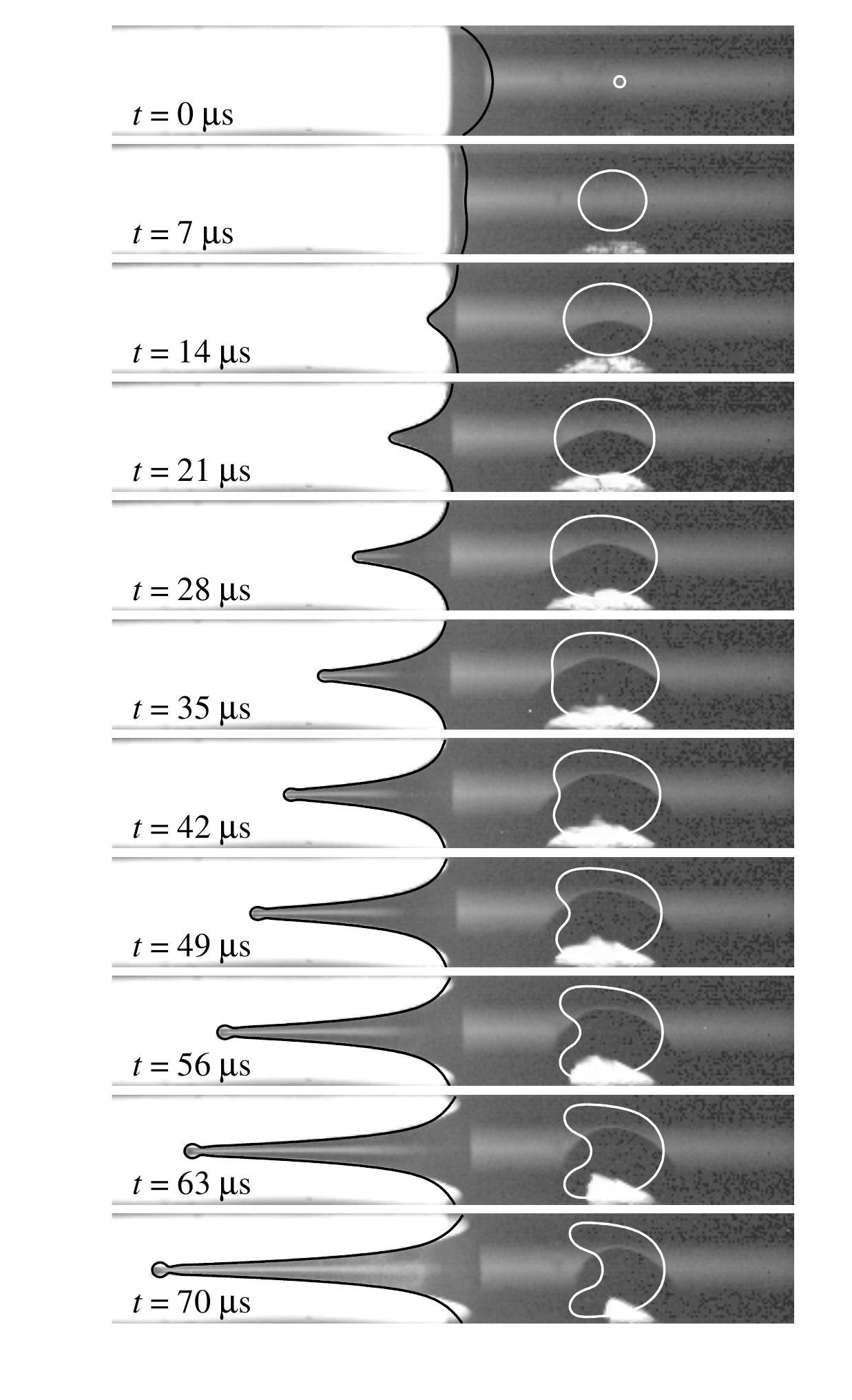}}
  \caption{Jet formation in experiment (background images) and simulation (black and white solid lines). Absorbed laser energy was $365~\mathrm{\mu J}$, distance between laser focus and meniscus was $600~\mathrm{\mu m}$, the tube radius is $250~\mathrm{\mu m}$. Pressure amplitude used in the simulation was $1581~\mathrm{bar}$, pressure duration was $50~\mathrm{ns}$, and the initial bubble radius was $25~\mathrm{\mu m}$.}
  \label{fig:expsim_free_surface}
\end{figure}

In this paper, we present numerical simulations which accurately reproduce the evolution of the shape and the velocity of the jets observed in the experiments described in \cite{Tagawa2012}. We use axisymmetric boundary integral simulations where we model the effect of the traveling pressure wave by applying a short pressure pulse on a bubble with a constant amplitude $\Delta p$ and a (short) duration $\Delta t$ such that the resulting impulse per unit area $\Delta p \Delta t$ is of the order of $10~\mathrm{Pa\cdot s}$. Figure~\ref{fig:expsim_free_surface} shows the formation of a jet in the experiment together with a result from our boundary integral simulations.
We provide a theoretical analysis that gives the correct scaling for the jet speed as a function of the contact angle, bubble distance and pressure pulse.

Jets that are formed from a meniscus have been studied in several cases with different sources for the driving pressure: In \cite{Duchemin2002}, the pressure was provided by surface tension, and \cite{Antkowiak2007} created a pressure pulse by impacting a test tube on a hard surface. In \cite{Lorenceau2002} and \cite{Bergmann2008} the driving was provided by hydrostatic pressure. In all these cases the scale of the jets is much larger and the velocities are at least an order of magnitude smaller than those that we study here.

The formation of the jet is different from the jet that follows from the collapse of a liquid void \citep{Longuet-Higgins1995,Hogrefe1998,Bergmann2006,Gekle2009}, where the jet initiates from a geometric singularity. In that case the size and the initial speed of the jet following the collapse of a cavity is therefore not set by the typical size and velocity of the experiment. Neither does the theory of a hyperbolic jet \citep{Longuet-Higgins1983} apply here. The main ingredient of the latter is a hyperbolic radial inflow from infinity, which in our setup is impossible due to the confinement of the tube. In fact, the jet that we study scales with the size of the capillary, and the speed is controlled by a combination of driving and geometry.

The paper is organized as follows. First we introduce the numerical method in \S\,\ref{sec:numset}, including a discussion of the geometrical setup and a discussion of how the simulations connect to the experiments of \cite{Tagawa2012}. Subsequently, we present the results from the numerical simulations and compare these to the experiments in \S\,\ref{sec:numres}, discussing the influence of the various parameters on the maximum velocity of the jet. In \S\,\ref{sec:model} we then derive an analytical model that predicts the achieved velocities, and we end with the conclusions and discussion (\S\,\ref{sec:concl}).

\section{Numerical setup}\label{sec:numset}
We perform numerical simulations using a boundary integral code, as described in~\cite{Oguz1993}, \cite{Bergmann2009}, \cite{Gekle2010b} and \cite{Gekle2010}. Here, we repeat the basic principles and methods, and elaborate on the parts that are specific for our case.

We approximate the flow in our system to be incompressible, inviscid and irrotational, so that we can describe the velocity field $\textbf{v}$ as the gradient of a potential $\phi$
\begin{equation}
    \textbf{v}=\nabla\phi
\end{equation}
which satisfies the Laplace equation
\begin{equation}
    \nabla^2\phi=0.
\end{equation}
Using Green's identity, the potential at any point inside the liquid domain can be described by an integral over the boundary containing $\phi$ and $\phi_n$, where $\phi_n$ is the spatial derivative of $\phi$ in the direction normal to the boundary. The system can be solved if at every point on the boundary either $\phi$ or $\phi_n$ is known. Solving the system is greatly simplified by imposing axial symmetry, reducing the surface integrals to line integrals. This simplification is justified by the axial symmetry observed in the experiments. On stationary solid boundaries we have $\phi_n=0$, and on the free surface we know the potential after time-integrating the unsteady Bernoulli equation
\begin{equation}
    \frac{\partial\phi}{\partial t} = -\frac{1}{2}|\nabla\phi|^2 - \frac{\Delta p + \kappa\sigma}{\rho} - gz
\end{equation}
with $\Delta p=p_g-p_a$ the pressure of the ambient vapour minus the atmospheric pressure, $\kappa$ the curvature, $\sigma$ the surface tension, $\rho$ the liquid density, $g$ the gravitational acceleration and $z$ the vertical coordinate. Due to the size and the time scale of the experiment, the gravitational component $gz$ can be neglected.
After solving the boundary integral equation, $\phi$ and $\phi_n$ are known everywhere on the boundary, and the new position of the free surface can be achieved by time-integrating the kinematic boundary condition
\begin{equation}
    \frac{\mathrm{d}\textbf{r}}{\mathrm{d}t}=\nabla \phi.
\end{equation}

Due to the absence of viscosity, some form of surface smoothing is necessary to keep the simulation stable. We use the node-shifting technique described by \cite{Oguz1990}, according to which new nodes are placed half way between all existing nodes, after which the original nodes are removed. This method effectively removes instabilities that are related to the node spacing everywhere on the free surface, except at the node on the symmetry axis because this node cannot be shifted or removed. We found that in our situation this node was subject to these instabilities, and therefore we applied an additional smoothing to it. This was done with the help of quadratic extrapolation of the position and the potential, using the two nodes next to the axis of symmetry and the symmetry condition. We verified that the numerical solutions were not sensitive to the amount of smoothing that we applied.

Because we are investigating a liquid inside a capillary, we have to take into account a moving contact line. The node that connects the liquid surface to the solid boundary can be considered both as part of the free surface and of the capillary wall. In solving the boundary integral equation, we treat this connecting node as a the latter, and impose $\phi_n=0$ on it. Implementation of the actual moving contact line with a dynamic contact angle as described by \cite{Voinov1976} would not be appropriate here since this is based on a balance between surface tension and viscosity, while our simulations are inviscid. Instead, we calculate the new position of the connecting node by extrapolating the nodes next to the connecting node. This method is similar to the one used by \cite{Oguz1993}, where the node connecting to the solid was displaced so that the contact angle remained at $90^\circ$. The exact implementation turns out only to have a non-negligible effect only close to the contact point and not to be important for the development of the jet. Comparing the extrapolation method and fixed contact angles between $60^\circ$ and $120^\circ$ resulted in less than $1\%$ variation of the maximum jet velocity.

\subsection{Initial condition}\label{sec:initial_cond}
\begin{figure}
  \centerline{\includegraphics[width=10cm]{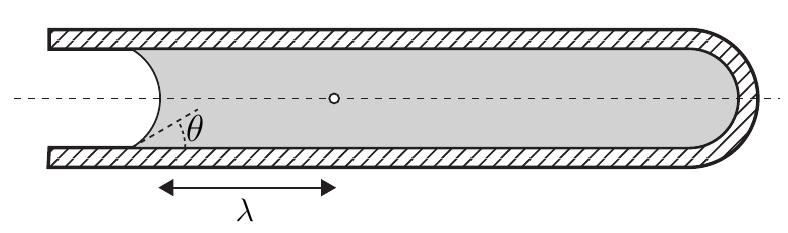}}
  \caption{The axisymmetric numerical setup. A tube of radius $R_t$ is filled with liquid. The liquid-air interface has the shape of a spherical cap, with a contact angle $\theta$. We position a bubble with a radius of $1/10$ of the tube radius at a distance $\lambda$ from the meniscus. The dashed line in the center represents the axis of symmetry.}
  \label{fig:numerical_setup}
\end{figure}

The computational domain that we use is closed at one side in the shape of a half-sphere; the free surface is at the opposite side of the tube (see figure~\ref{fig:numerical_setup}). Because the diameter of the tube is much smaller than the capillary length, gravity can be safely neglected, and the free surface adopts the shape of a spherical cap. The initial shape of the free surface in the simulations can therefore be defined using only the static contact angle $\theta$. Note that the contact angle $\theta$ only serves as an initial condition, and we do not impose a dynamic contact angle. A bubble with a radius $1/10$ of the tube radius is positioned at a distance $\lambda$ from the free surface. The distance of the bubble from the closed end of the tube does not have an influence on the simulations, as long as it is a few tube radii or more.

\subsection{Pressure wave model}
In the experiment, a pressure wave is created by vaporizing a small amount of liquid with a laser pulse. This abrupt vaporization is responsible for a very large increase in the pressure in a small volume, which results in a pressure wave that travels through the tube and reflects on the free surface \citep{Tagawa2012}. As argued in that paper, the initial velocity $V_0$ of the free surface is connected to the pressure wave strength $\Delta p\approx\thalf\rho c V_0$, where $c$ is the speed of sound in the liquid. E.g., for $V_0\sim10~\mathrm{m/s}$ and $c=1497~\mathrm{m/s}$ we find $\Delta p\sim75$ bar.

\begin{figure}
  \centerline{\includegraphics[width=8cm]{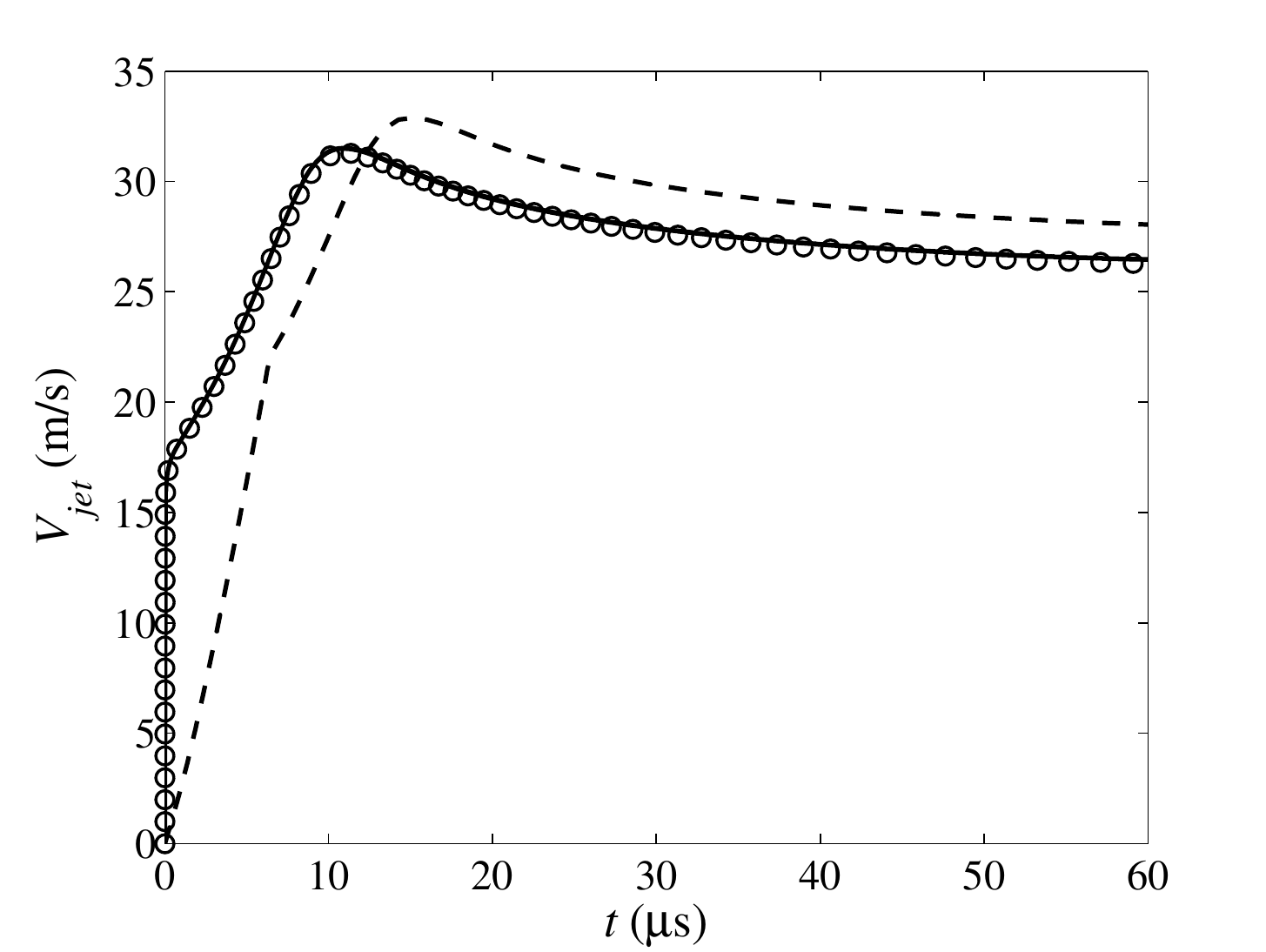}}
  \caption{The velocity of the tip of the jet as a function of time for different values of $\Delta t$. The product $\Delta t\Delta p$ is kept constant at $15.20~\mathrm{Pa\cdot s}$, showing that the velocity of the jet only depends on this product when $\Delta t$ is small enough. The results for $\Delta t=25~\mathrm{ns}$ (circles, $\Delta p=6080~\mathrm{bar}$) and $\Delta t=50~\mathrm{ns}$ (solid line, $\Delta p=3040~\mathrm{bar}$) overlap, whereas a significantly larger duration of the pressure pulse $\Delta t=6.4~\mathrm{\mu s}$ (dashed line, $\Delta p=23.75~\mathrm{bar}$) results in a different velocity and a different evolution of the velocity.}
  \label{fig:shock_wave_model}
\end{figure}

The reflections of the pressure wave on the free surface and the wall of the tube ultimately result in a pressure gradient between the vapor bubble and the free surface, so that the entire liquid volume in between will start to move. Starting from this very early point in time, the dynamics of the system is expected to be well described by the potential flow boundary integral model employed in this paper. In the simulation, we model this pressure wave by applying a pressure pulse on the bubble \citep{Ory2000}. This pressure pulse has a typical amplitude $\Delta p$ and a duration $\Delta t$. Figure \ref{fig:shock_wave_model} shows that if $\Delta t$ is small enough, the only relevant value is the product $\Delta p\Delta t$ (which is in the order of $10~\mathrm{Pa\dot s}$) and the pressure pulse can be assumed to be instantaneous. For simplicity, we will keep $\Delta t$ at $50~\mathrm{ns}$, and vary only the pressure amplitude. With this choice, the pressure amplitude easily reaches values in the order of $10^3~\mathrm{bar}$, i.e., in excess of the critical pressure of water ($\approx 220~\mathrm{bar}$). This does not need to worry us too much, since not $\Delta p$ itself but the product $\Delta p \Delta t$ determines the course of events. After the pressure pulse, the pressure inside the bubble is set to $0$ to account for the rapid condensation of the vapour in the bubble as was also done by \cite{Ory2000}. A more sophisticated model where the heat exchange is taken into account for the growth and collapse of a vapour bubble created by a laser pulse can be found in~\cite{Sun2009}. In our case, applying a perfect gas law as well as incorporating heat transfer only resulted in marginal differences in the jet. We therefore use a simpler model here, which minimizes the number of unknown adjustable parameters. The good agreement between the simulations and the experiments convinces us that our model is accurate enough to describe the physics that create the jets observed in experiments.

\section{Numerical results}\label{sec:numres}
We will show here the results from the numerical simulations, compare them to the experiments of \cite{Tagawa2012}, and discuss the influence of various parameters. More specifically, we will first compare the time-evolution of the shape and speed of the jet from the numerical simulations with the experiments. After this, we will focus on the maximum velocity of the jet by investigating the influence of the contact angle, surface tension, pressure impulse, tube radius, and bubble distance, and discuss these results in the context of the experiments.

In figure \ref{fig:expsim_free_surface} we find 11 snapshots from the experiment, overlaid with the corresponding BI results. At $t=0$ the laser pulse is absorbed which is modelled in the simulation as a bubble pressurized to $1581~\mathrm{bar}$ for a time span of $50~\mathrm{ns}$. The initially concave interface starts to move instantly and goes through an almost flat stage at $t=7~\mathrm{\mu s}$ to the development of a jet ($t=14-56~\mathrm{\mu s}$). The maximum jet speed is reached between the second and the third frame at $t=10~\mathrm{\mu s}$, when the jet just starts to form.
The position of the tip of the jet is fully reproduced, as well as the overall shape of the jet. The bubble in the experiment is created near the wall of the capillary, resulting in a difference in shape, but its size is reproduced by the simulations at least up to $t=35~\mathrm{\mu s}$, as can be appreciated by considering volume conservation in the system. Because the full free surface of the experimental jet is reproduced by the simulation up to $t=35~\mathrm{\mu s}$, we know that the volume of the bubble in the simulation is also the same as in the experiment. After this there is a slight difference in the collapse of the bubble, which results in a small difference at the base of the jet where the free surface connects to the wall of the capillary. This however has no significant effect on the part of the jet that is further away from the contact point.

\subsection{Jet and bubble shape}
\begin{figure}
  \centerline{\includegraphics[width=8cm]{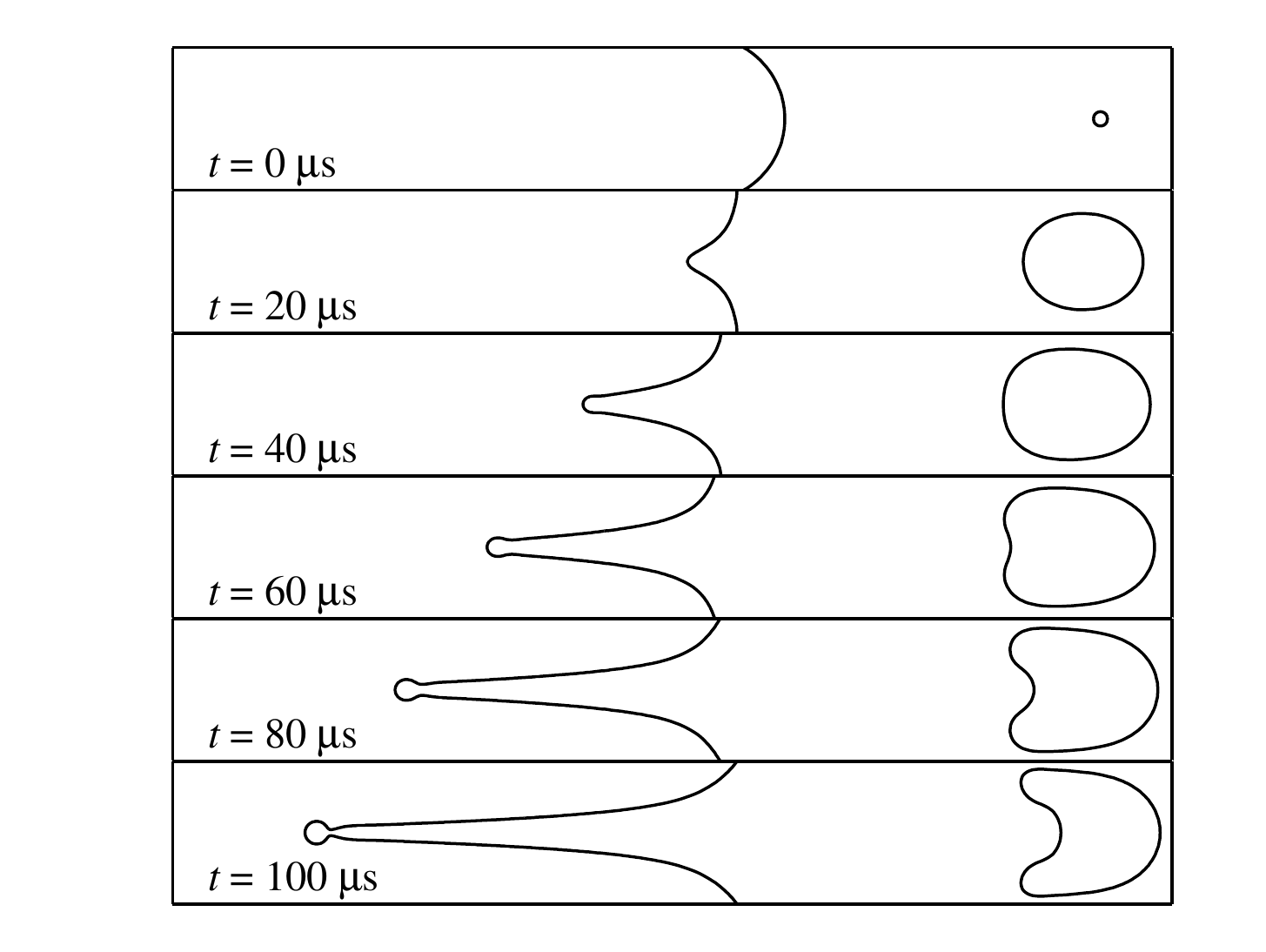}}
  \caption{The evolution of the jet and the bubble from a boundary integral simulation. Parameters: $R_{tube}=250~\mathrm{\mu m}$, $\Delta p=2027~\mathrm{bar}$, $\Delta t=50~\mathrm{ns}$, $\theta=30~\mathrm{degrees}$, $\lambda=1106~\mathrm{\mu m}$.}
  \label{fig:jet_shapes}
\end{figure}

Figure~\ref{fig:jet_shapes} shows how the jet and the bubble develop in time. The maximum speed of the jet is approximately $30~\mathrm{m/s}$, which is reached around $20~\mathrm{\mu s}$ after the pressure pulse. The jet has a diameter which is about $1/10$ of the diameter of the tube; this holds for all tube diameters that we tested. The bubble initially grows spherically but, due to confinement and asymmetry, it later takes on an elongated shape and grows more towards the free surface. The right side of the bubble is almost stationary, also during the collapse where a thick jet is formed reminiscent of the collapse of a bubble near a free surface or a solid boundary \citep{Blake1981,Lindau2003}.

\subsection{Jet velocity and velocity field}
\begin{figure}
  \centerline{\includegraphics[width=8cm]{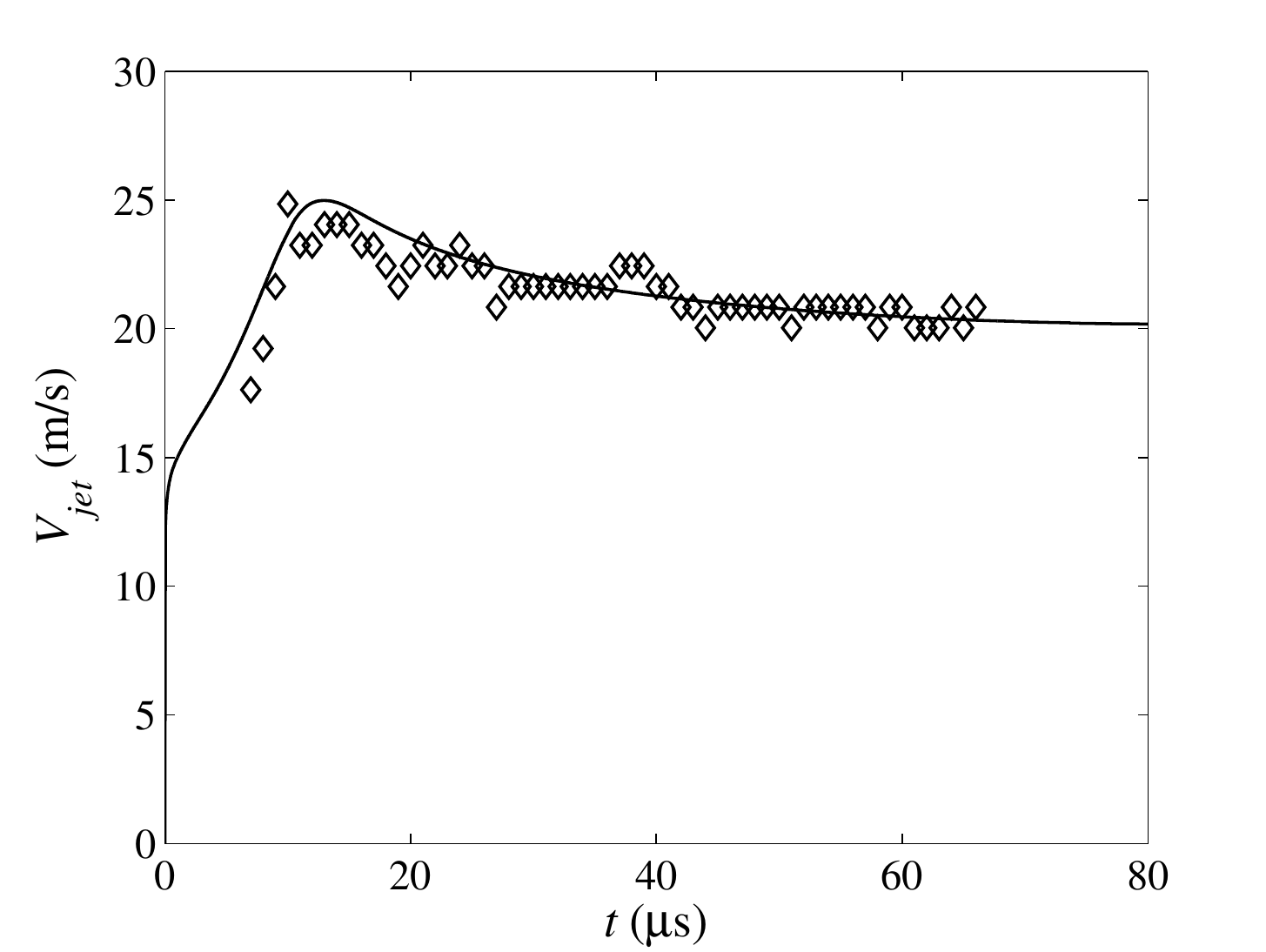}}
  \caption{The velocity of the tip of the jet as a function of time for both the simulation (solid line) and experiments (diamonds). After an almost instantaneous acceleration to $13~\mathrm{m/s}$ during the $50~\mathrm{ns}$ pressure pulse, the tip is further accelerated by the focusing geometry to about $25~\mathrm{m/s}$. Experimental conditions and numerical settings are the same as in Fig. \ref{fig:expsim_free_surface}.}
  \label{fig:jet_speed}
\end{figure}

In this paper we focus on the jet velocity which can directly be compared with the experimental measurements. We define the jet velocity as the velocity component parallel to the tube axis at the tip of the jet (in figure~\ref{fig:jet_shapes} to the left). Figure~\ref{fig:jet_speed} shows numerical results together with experimental results on how the jet velocity evolves in time. There are two acceleration mechanisms: First, driven by the very short pressure pulse, a speed of about $13~\mathrm{m/s}$ is reached almost instantaneously. After this there is no more driving, but the focusing of the flow accounts for a further acceleration of the jet which reaches a maximum velocity of about $25~\mathrm{m/s}$. Deceleration is accounted for by surface tension (see \S\,\ref{sec:surface_tension})  and the collapsing bubble.

\begin{figure}
  \centerline{\includegraphics[width=12cm]{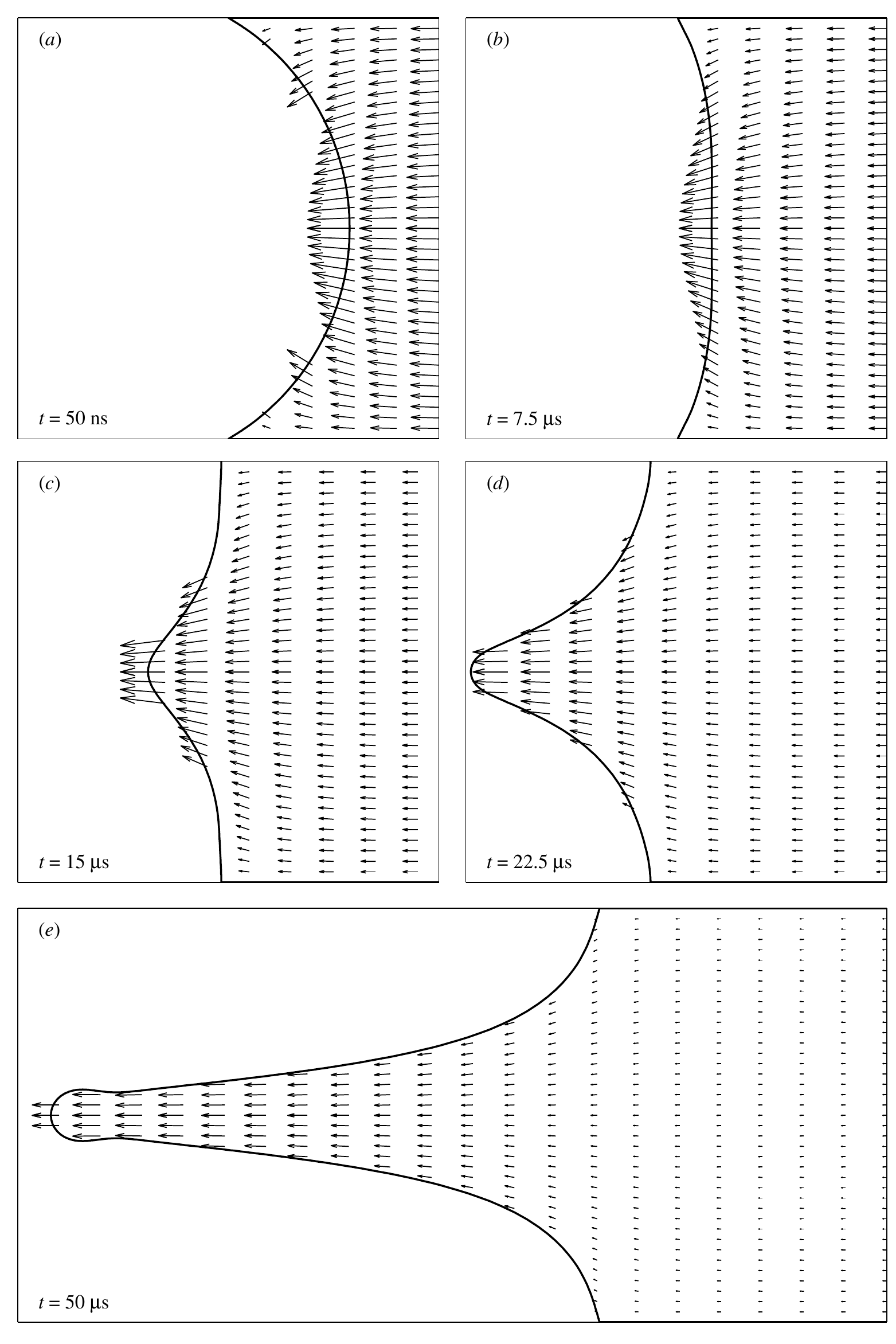}}
  \caption{The velocity field during the formation of the jet. (\textit{a}) is right after the pressure pulse, the free surface still has a spherical cap shape. In (\textit{a}), (\textit{b}) and (\textit{c}), focusing of the flow can be seen. (\textit{d}) and (\textit{e}) clearly show the stretching of the jet: The largest velocity is in the tip of the jet, and gradually decreases towards the base of the jet.}
  \label{fig:velocity_field}
\end{figure}

Figure~\ref{fig:velocity_field} shows the velocity field in the liquid during jet formation. In figure~\ref{fig:velocity_field}(\textit{a}) the interface has not moved significantly due to the small time interval, but it clearly shows how the interface is responsible for the focusing of the flow. In figure~\ref{fig:velocity_field}(\textit{b}) we see that although the surface in the center is approximately flat, the velocity still has a focusing profile. Indeed, at $t=7.5~\mathrm{\mu s}$ the jet is still accelerating. Only after about $15~\mathrm{\mu s}$ (see figure~\ref{fig:velocity_field}(\textit{c})), there is no more focusing of the flow. Stretching of the jet is visible in figure~\ref{fig:velocity_field}(\textit{d}) and \ref{fig:velocity_field}(\textit{e}) where the velocity of the tip of the jet is larger than the velocity at the base, with the consequence that the jet is becomes thinner while increasing in length \citep{Eggers2008}.

\subsection{Contact angle}
Now, how does the jet speed depend on the contact angle? As we explained in \S\,\ref{sec:initial_cond} the meniscus initially has the shape of a spherical cap. The meniscus thus has a well-defined radius of curvature, which depends on the contact angle $\theta$. A contact angle of $90^\circ$ results in a flat interface (no curvature), and a contact angle of $0^\circ$ gives a radius of curvature equal to the inner radius of the capillary. A smaller contact angle increases the curvature of the free surface, and therefore increases the focusing of the flow.

\begin{figure}
  \centerline{\includegraphics[width=8cm]{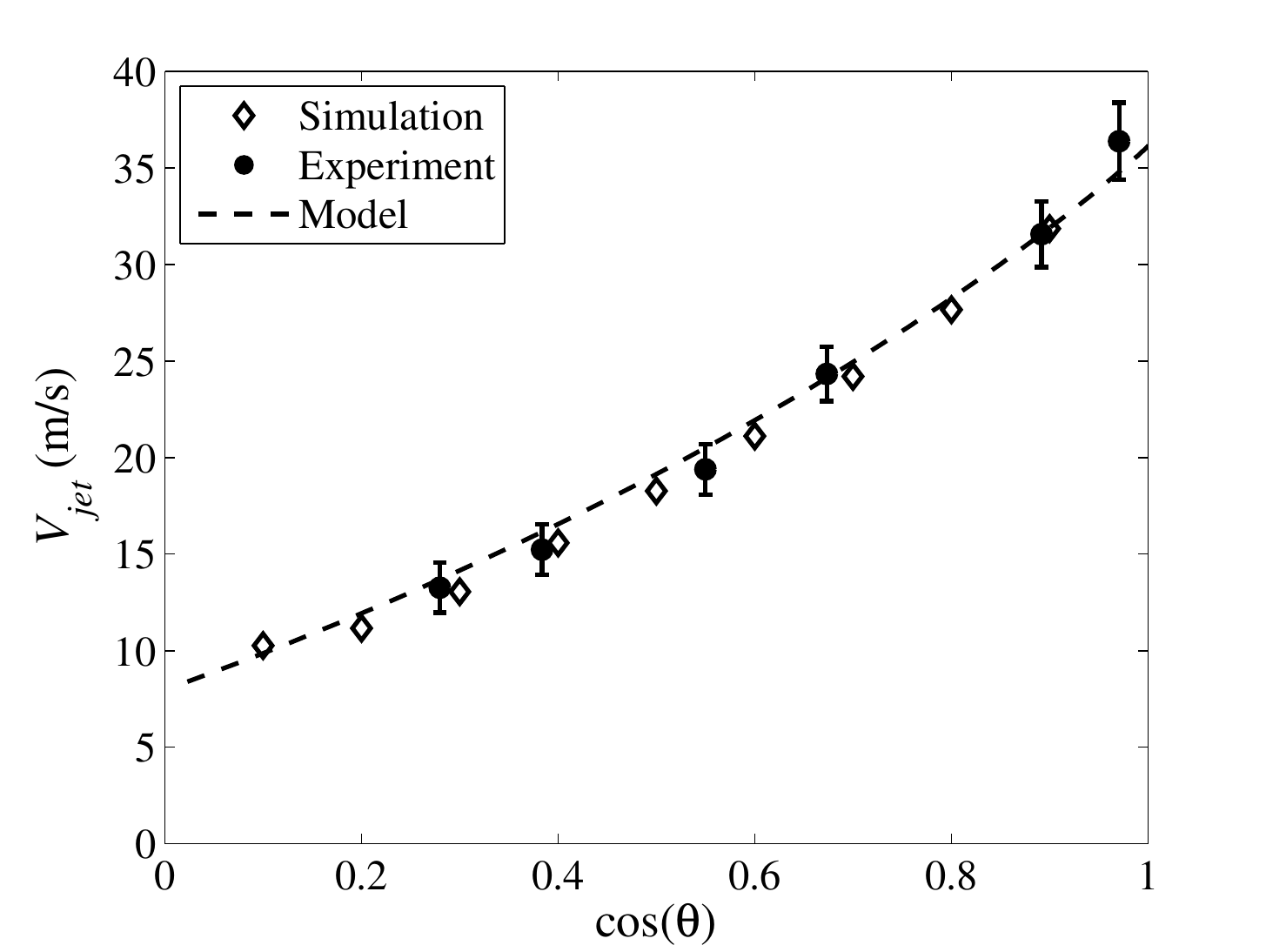}}
  \caption{The influence of the contact angle on the jet velocity. Tube radius is $250~\mathrm{\mu m}$, and $\lambda=460~\mathrm{\mu m}$. The numerical results are obtained with $\Delta p=1647~\mathrm{bar}$, the absorbed laser energy in the experimental data was $458~\mathrm{\mu J}$. The dashed line represents the model~(\ref{model_correct_Vmax}), with $\alpha=0.44$, $\beta=1.33$, and $h_0=0.26$.}
  \label{fig:theta}
\end{figure}

In the experiments, $\theta$ can be measured directly from images of the static meniscus, so we can directly compare the influence of the contact angle in experiments and simulations. Figure~\ref{fig:theta} shows that the jet velocity clearly increases with $\cos\theta$. A more careful look reveals that the dependence on the contact angle is a bit stronger than linear, both in the experiment and the computation. There is a non-linear dependence on $\cos\theta$ because the curved interface increases the velocity in two stadia of the jet formation: During the pressure pulse and during the flow focusing. In \S\,\ref{sec:model} we will provide a detailed analysis of this dependence on $\theta$. The absorbed laser energy in the experiments in figure~\ref{fig:theta} was $458~\mathrm{\mu J}$, the applied pressure in the simulations was $1674~\mathrm{bar}$.

\subsection{Surface tension}\label{sec:surface_tension}
In order to study the effect of surface tension in isolation, we turn off the collapse mechanism of the bubble by setting the pressure in it to atmospheric after the initial pressure pulse. In this case the bubble keeps growing until it would be ultimately restrained by surface tension over a much longer time scale than we consider.

\begin{figure}
  \centerline{\includegraphics[width=8cm]{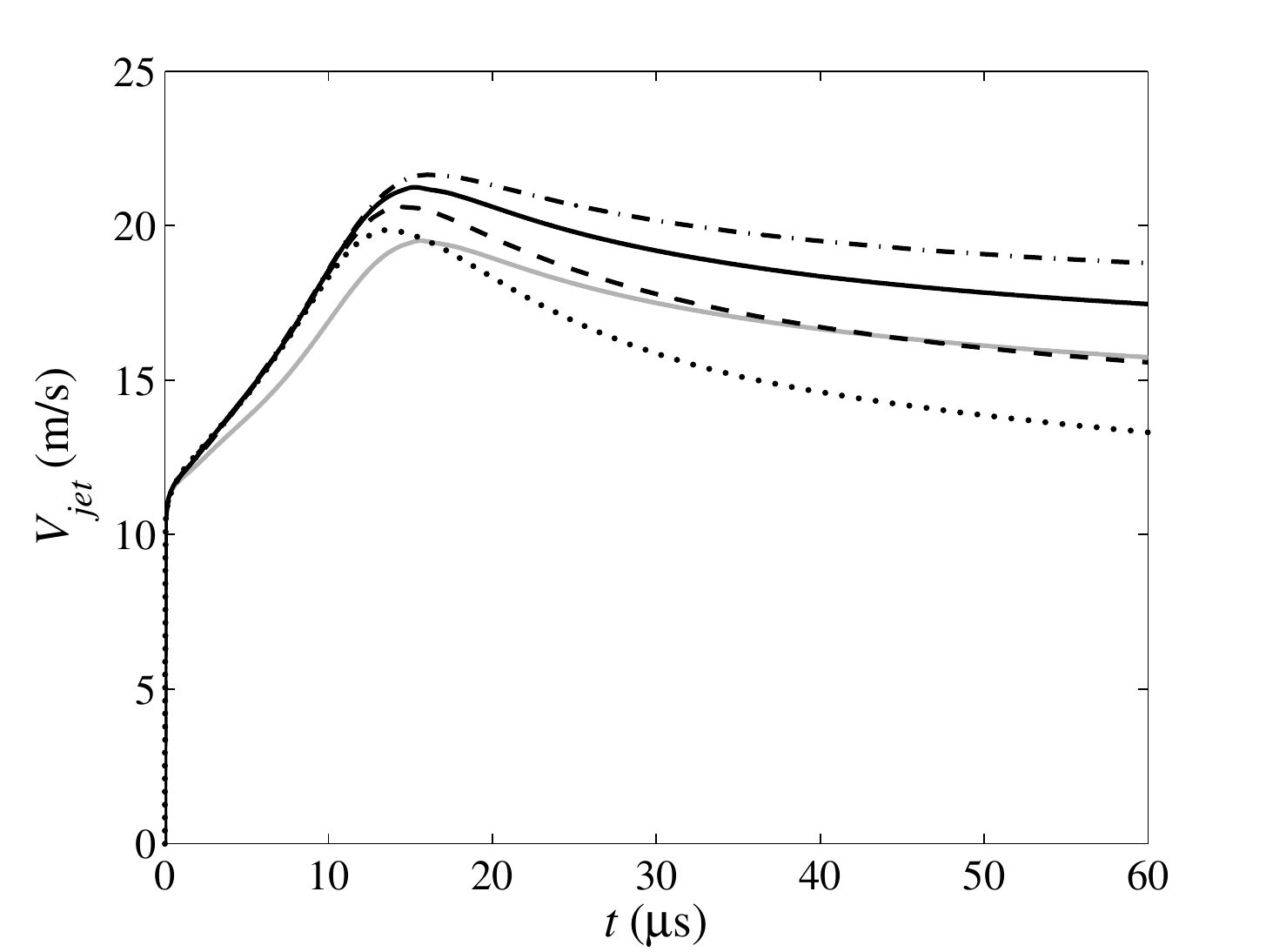}}
  \caption{Jet speed versus time for different surface tensions. The solid line has the surface tension of water ($\sigma=72.8~\mathrm{mN/m}$), the other values for the surface tension are $\sigma=35~\mathrm{mN/m}$ (dash-dotted line), $\sigma=150~\mathrm{mN/m}$ (dashed line) and $\sigma=300~\mathrm{mN/m}$ (dotted line). The collapse of the bubble was turned off in these simulations to isolate the effect of surface tension. Tube radius is $250~\mathrm{\mu m}$, $\lambda=1106~\mathrm{\mu m}$, and $\Delta p=2027~\mathrm{bar}$. For reference, the evolution of the jet velocity with bubble collapse is represented by the gray solid line, with the same parameters and $\sigma=72.8~\mathrm{mN/m}$.}
  \label{fig:surface_tension}
\end{figure}

Figure~\ref{fig:surface_tension} shows the development of the jet velocity for four different values of the surface tension. Clearly, the acceleration phase is dominated by inertia, as there is almost no difference in the acceleration while there is an order of magnitude difference in the surface tension. Only when the jet reaches its maximum velocity and during deceleration surface tension starts to play a role. This stands to reason, because the only decelerating mechanism in this case is  surface tension, counteracting the increase of surface area caused by the jet.

We can calculate the order of magnitude of the expected decrease in velocity $V_s$ using an argument similar to the Taylor-Culick velocity of a liquid sheet \citep{Taylor1959,Culick1960,Gordillo2010}:
Using the jet radius $R_j$ as the relevant length scale, we can balance the kinetic energy per unit length
$E_k \sim \thalf\pi \rho R_j^2 V_s^2$
with the surface energy of the jet per unit length $E_s \sim 2\pi \sigma R_j$
and arrive at a velocity
\begin{equation}
    V_s \sim \left(\frac{4\sigma}{\rho R_j}\right)^{1/2}
\end{equation}
Using $R_j\approx25~\mathrm{\mu m}$ and the material properties of water gives $V_s \sim 3.4~\mathrm{m/s}$, which is indeed similar to the decrease in velocity presented by the solid line in Figure~\ref{fig:surface_tension}.

\subsection{Pressure impulse}
\begin{figure}
  \centerline{\includegraphics[width=\textwidth]{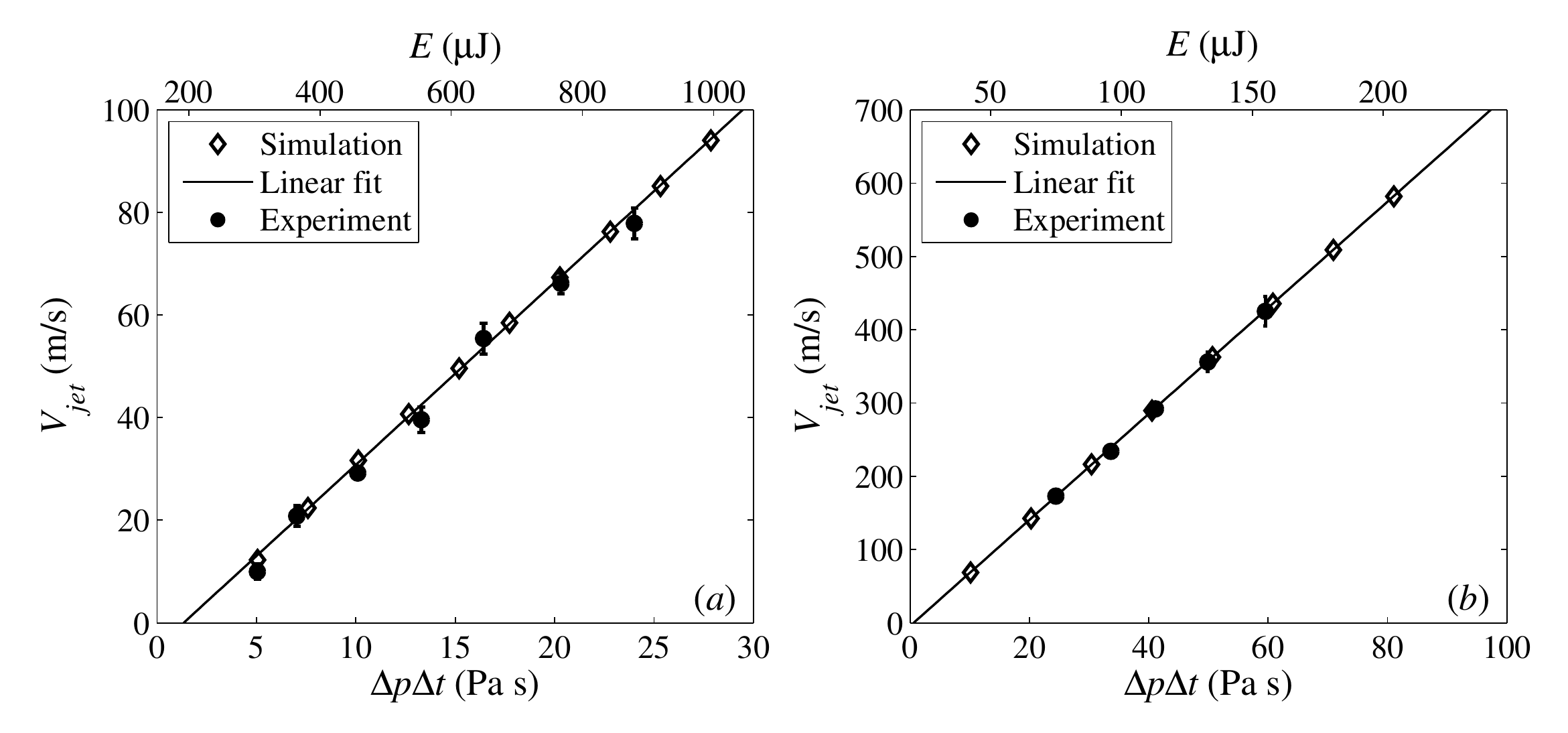}}
  \caption{The influence of the pressure on the jet velocity, compared to experiments for two different tube radii. (\textit{a}): Tube radius $250~\mathrm{\mu m}$, $\Delta t=50~\mathrm{ns}$, $\theta=30~\mathrm{degrees}$ and $\lambda=600~\mathrm{\mu m}$. (\textit{b}): Tube radius $25~\mathrm{\mu m}$, $\Delta t=5~\mathrm{ns}$, $\theta=30~\mathrm{degrees}$ and $\lambda=400~\mathrm{\mu m}$. Experimental data were converted from the absorbed laser energy to pressure impulse by a fitting routine; the energy is indicated at the top axis.}
  \label{fig:pressure}
\end{figure}

The magnitude of the pressure impulse $\Delta p \Delta t$  is the only parameter that cannot be directly related to the experiment. We expect that the absorbed laser energy is the experimental parameter that is most directly related to it. In the experiments, a linear relation between the energy and the velocity of the jet is found. As can be seen in figure~\ref{fig:pressure}, in the numerical simulation the jet velocity also depends linearly on the pressure pulse. From these observations we conclude that there is a linear relation between the absorbed laser energy ($E$ in Joules) as measured in the experiments and the pressure impulse that we apply in the simulations of figure~\ref{fig:pressure}:
\begin{equation}
    \Delta p\Delta t \approx 3.30\cdot10^4~\mathrm{s\,m^{-3}}\,E - 5.0 ~\mathrm{Pa\cdot s}
    \label{eq:pressure250}
\end{equation}
for $R_{tube}=250~\mathrm{\mu m}$, and
\begin{equation}
    \Delta p\Delta t \approx 4.39\cdot10^5~\mathrm{s\,m^{-3}}\,E - 8.5 ~\mathrm{Pa\cdot s}
\end{equation}
for $R_{tube}=25~\mathrm{\mu m}$. The prefactor is one order of magnitude larger when the tube radius is one order smaller, while the threshold value is of the same order. More about the influence of the tube radius can be found in \S\,\ref{sec:tube_radius}.

In both the simulation and the experiment, there is an apparent threshold value for the energy or pressure below which we can not observe a well-defined jet. In the experiments, the main reason for this would be that a large amount of the laser energy is lost in heating up the fluid before a bubble can be created, as was shown by~\cite{Sun2009}. In the simulations the only cause for the threshold lies in the surface tension that prevents the formation of a jet, and the vacuum inside the bubble after the pressure pulse is applied. Once the kinetic energy is much larger than the surface energy related to the formation of the jet and the potential energy related to the size of the vacuum bubble, a jet can be formed. The zero value of $E$ extrapolated on the upper horizontal scale in figure~\ref{fig:pressure} lies considerably to the left of the zero value of the lower impulse scale, which implies that the experimental threshold, due to thermal and other losses, is significantly higher than the numerical one.

\subsection{Tube radius}\label{sec:tube_radius}
\begin{figure}
  \centerline{\includegraphics[width=8cm]{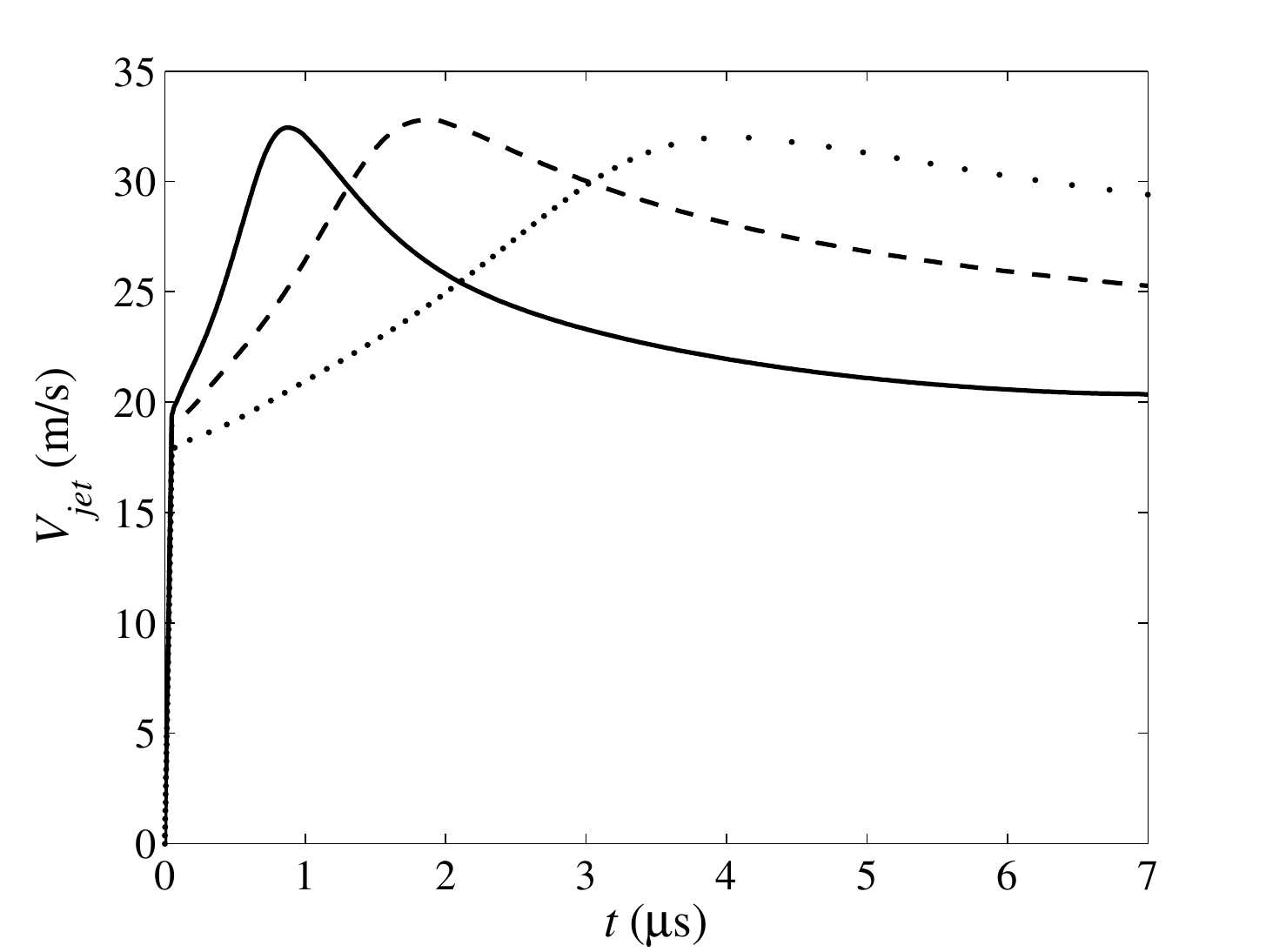}}
  \caption{The influence of the tube radius on the jet velocity. A smaller radius increases the acceleration due to flow focusing, but decreases the duration of the acceleration, resulting in approximately the same maximum jet velocity. Tube radii in this figure are $25~\mathrm{\mu m}$ (solid line), $50~\mathrm{\mu m}$ (dashed line) and $100~\mathrm{\mu m}$ (dotted line). The other parameters are the same for all three simulations: $\Delta p=3040~\mathrm{bar}$, $\Delta t=50~\mathrm{ns}$, $\theta=30~\mathrm{degrees}$ and $\lambda=1250~\mathrm{\mu m}$.}
  \label{fig:diameter}
\end{figure}

In the experiments, there is a clear dependence of the jet velocity on the tube radius $R_t$: smaller tubes create faster jets with the same absorbed laser energy, approximately following the relation $V_{jet}\propto1/R_t$. One naively would argue that a smaller tube will provide a stronger curved free surface, and therefore the stronger focusing will result in a faster jet. In contrast, Figure~\ref{fig:diameter} shows that the maximum jet speed that is obtained in the simulations at most only shows a very weak dependence on the tube radius (if there is any at all). However, a different effect caused by the tube radius is very clear: The acceleration is much larger for smaller tubes, so that the maximum velocity is reached earlier. Indeed, a smaller tube provides a higher curvature of the free surface, so the acceleration due to flow focusing is larger. The maximum velocity, however, is a combination of the magnitude and the duration of the acceleration, which both depend on the tube radius. The simulations show that these two parameters cancel each other if we only change the tube radius: larger tubes have less acceleration due to focusing of the flow, but the acceleration persists for a longer time, as can be seen clearly in figure~\ref{fig:diameter}. This is consistent with dimensional analysis, provided that the relevant time and velocity scales are taken to be $R_t/V_0$ and $V_0 = \Delta p \Delta t/(\rho \lambda)$, which will be motivated and discussed in \S\,\ref{sec:model}. This scaling will make the curves collapse approximately up to the maximum, confirming that the tube radius does not influence the jet velocity. After the maximum there is no collapse due to the influence of surface tension (cf. \S\,\ref{sec:surface_tension}).

Although our simulations and analysis show that the jet velocity is independent of the tube radius, the remaining question is why there \emph{is} such a strong dependence on the tube radius in the experiments. The most plausible explanation is that for a fixed absorbed energy, the generated pressure has a strong dependence on the tube radius. The reason for this could be found in the volume $V_e$ that is exposed to the laser energy $E$. Based on dimensional analysis the produced pressure can be expected to scale as
\begin{equation}
    \Delta p\propto\frac{E}{V_e}.
\end{equation}
A smaller tube would result in a smaller volume that is exposed to the laser energy. This then would account for the dependence of the jet velocity on the tube radius that was observed in the experiments.

\subsection{Bubble distance from the free surface}
In experiments, the most direct measurable parameter is the distance between the meniscus and the bubble. This makes it an excellent candidate to compare with numerical simulations. We define the distance $\lambda$ as the distance between the center of the bubble and the point on the meniscus that is on the axis of symmetry (see figure~\ref{fig:numerical_setup}). Due to the axisymmetry of the numerical simulations, the bubble is always in the center of the tube. In experiments however, the bubble is usually created near the wall of the tube, due to the characteristics of the absorption of the laser light in the liquid. The difference between these different bubble positions can be neglected when the distance between the bubble and the meniscus is large compared to the radius of the tube ($\lambda/R_t\gg1$).

\begin{figure}
  \centerline{\includegraphics[width=8cm]{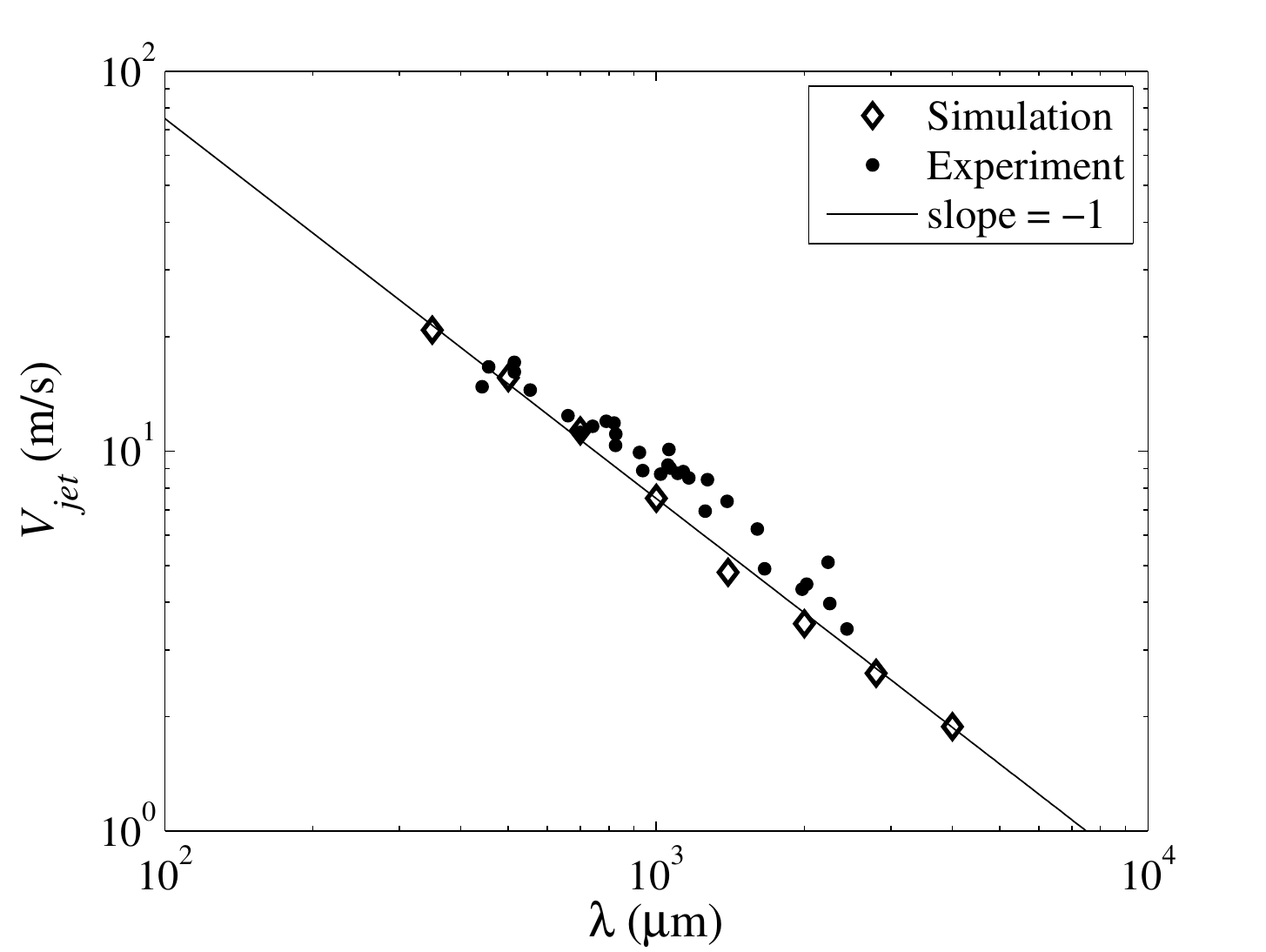}}
  \caption{The influence of the distance between the bubble and the meniscus on the jet velocity. Tube radius is $250~\mathrm{\mu m}$ and $\theta=25^\circ$. The numerical results are obtained with $\Delta p=1013~\mathrm{bar}$, directly related with equation~(\ref{eq:pressure250}) to the absorbed laser energy in the experimental data, which was $305~\mathrm{\mu J}$. A power-law fit (solid line) reveals a clear $1/\lambda$ dependence of the jet velocity.}
  \label{fig:lambda}
\end{figure}

Figure~\ref{fig:lambda} shows the jet velocity as a function of $\lambda$, together with the experimental measurements. There is a good agreement between experiments and simulations, and both show a clear $1/\lambda$ dependence for the jet velocity. Both the numerical and the experimental results are obtained with a tube radius of $250~\mathrm{\mu m}$ and a contact angle of 25~degrees. The absorbed laser energy in the experiments was $305~\mathrm{\mu J}$ and the pressure amplitude for the simulations, calculated using (\ref{eq:pressure250}), was $1013~\mathrm{bars}$.

\section{Theoretical approximation}\label{sec:model}
We will now try to analytically understand the mechanism of the jet formation and the achieved velocities by approaching the problem with a simplified model. The route through which the jet attains its maximum velocity can be split in two parts: the pressure impulse and the flow focusing. The effect of the former is determined by how the pressure wave, which gives an initial velocity to the liquid, is modelled in the (incompressible) simulations which will be discussed in \S\,\ref{sec:presspulse}. After the liquid is set into motion, the curved shape of the meniscus leads to a further acceleration of the liquid by focusing it in a fast thin jet (\S\,\ref{sec:flowfocus}). We first neglect the influence of the curvature of the free surface on the action of the pressure pulse to compute the maximum jet velocity (\S\,\ref{sec:maxvel}), and then revisit this approximation in \S\,\ref{sec:cormaxvel}.

\label{sec:flowfocus}
\label{sec:maxvel}
\label{sec:cormaxvel}

\subsection{The pressure pulse}\label{sec:presspulse}
We approximate the system during the pressure pulse as one-dimensional, so after neglecting viscosity and compressibility we can write the Euler equation as
\begin{equation}
    \frac{\partial u}{\partial t} = -\frac{1}{\rho}\frac{\partial p}{\partial z},
    \label{NavierStokes}
\end{equation}
where $u$ is the axial velocity of the liquid, and $t$ the time. Note that due to continuity in this one-dimensional system, the $\partial u/\partial z$ term in the material derivative equals $0$, so on the left hand side we only have $\partial u/\partial t$. The axial pressure gradient $\partial p/\partial z$ is given by the pressure in the bubble $\Delta p$ and the distance $\lambda$ between the bubble and the free surface:
\begin{equation}
    \frac{\partial p}{\partial z} = \frac{\Delta p}{\lambda}.
\end{equation}
The Laplace pressure jump on the free surface can be neglected because $\Delta p$ is very large compared to the typical pressure associated with surface tension.

We integrate \ref{NavierStokes} over the duration $\Delta t$ of the pressure pulse and obtain the velocity $V_0$ of the free surface after the pressure pulse \citep{Ory2000}:
\begin{equation}
    V_0 = \frac{\Delta p \Delta t}{\rho \lambda},
    \label{eq:V_0}
\end{equation}
where we assume $\lambda$ to be constant, which can be done if $\Delta t$ is small enough.

\subsection{The flow focusing}\label{sec:flowfocus}
\label{sec:flowFocusing}
After the pressure pulse there is no more driving of the flow, which means that all further acceleration is caused focusing. 

We will now give an analysis for the acceleration due to flow focusing based on continuity. Starting with a spherical surface with radius of curvature $R_c$ and velocity $V_0$ directed normal it, we keep the flow rate constant:
\begin{equation}
    V_0R_c^2 = (V_0+dV)(R_c-dR)^2,
    \label{eq:continuity}
\end{equation}
where $dV$ is a small increase in velocity due to a small decrease in radius $dR$. At leading order, $dR=V_0dt$, and (\ref{eq:continuity}) becomes:
\begin{equation}
    \frac{dV}{dt}=\frac{2V_0^2}{R_c}
\end{equation}
The radius of curvature can be expressed using the tube radius $R_t$ and the contact angle $\theta$ as $R_c=R_t/\cos\theta$, which then gives us the following expression for the acceleration:
\begin{equation}
    a=2V_0^2\frac{\cos\theta}{R_t}
\end{equation}
Note that the same scaling for the acceleration can be obtained by dimensional analysis using $V_0$ and $R_c$ as the relevant velocity and length scale respectively.

Clearly, smaller tubes have stronger focusing and therefore generate a larger acceleration. This, however, does not mean that the maximum velocity of the jet will be higher as well. To determine the increase in speed due to the flow focusing we have to find a time scale during which the fluid is accelerated. The focusing time scale $\Delta t_f$ is provided by the typical velocity $V_0$ (the velocity created by the pressure pulse) and the typical length scale $R_t$ (the radius of the tube):
\begin{equation}
    \Delta t_f = \frac{R_t}{V_0}.
\end{equation}
The increase in velocity due to flow focusing is then:
\begin{equation}
    \Delta V \sim a\Delta t_f = 2V_0\cos\theta,
\end{equation}
where it becomes clear that the increase in velocity due to flow focusing is independent of the tube radius.

\subsection{The maximum jet velocity}\label{sec:maxvel}
The maximum velocity reached by the jet is the sum of the velocity reached after the pressure pulse and the increase in velocity due to flow focusing:
\begin{equation}
    V_{max} = V_0+\Delta V = \frac{\Delta p\Delta t}{\rho\lambda}(1+\beta\cos\theta)
    \label{eq:V_max}
\end{equation}
with $\beta$ a proportionality factor which we expect to be of order unity. First of all, the proportionality to $\frac{\Delta p\Delta t}{\rho\lambda}$ is in excellent agreement with the results from the simulations shown in figures \ref{fig:pressure}, \ref{fig:lambda} and with the fact that $V_{max}$ does not depend on the tube radius (see figure \ref{fig:diameter}). To compare the dependence on the curvature of the meniscus, we now turn to figure~\ref{fig:jetvel_model}(\textit{a}). Here we compare the model with $\beta=2.0$ to the simulation data. The velocities are roughly reproduced, but it is clear that there is a dependence on the curvature for $V_0$, which is not accounted for by the model and the increase of $V_{max}$ is therefore not very accurately reproduced. Clearly, neglecting the curvature of the surface during the pressure pulse has been too bold an assumption.

\subsection{Correction for a pressure pulse on a curved interface}\label{sec:cormaxvel}
We will now apply a correction to the above derived model to account for the curved interface during the pressure pulse. The above derivation (\ref{eq:V_0}) gives the velocity in the bulk, far away from the the bubble and the free surface. We will use volume conservation and an approximate velocity distribution on the free surface to calculate the free surface velocity on the tube axis.

The first step will be to determine the velocity distribution on the free surface. Due to the short time scale and the magnitude of the pressure pulse, we can neglect the tangential velocity components. We therefore only take into account the velocity normal to the interface, and we can consider the free surface as an equipotential surface. Away from the free surface we expect the one-dimensional approximation to hold, resulting in a uniform axial velocity, so that we will have evenly spaced equipotential surfaces oriented perpendicular to the tube wall.

\begin{figure}
  \centerline{\includegraphics[width=6cm]{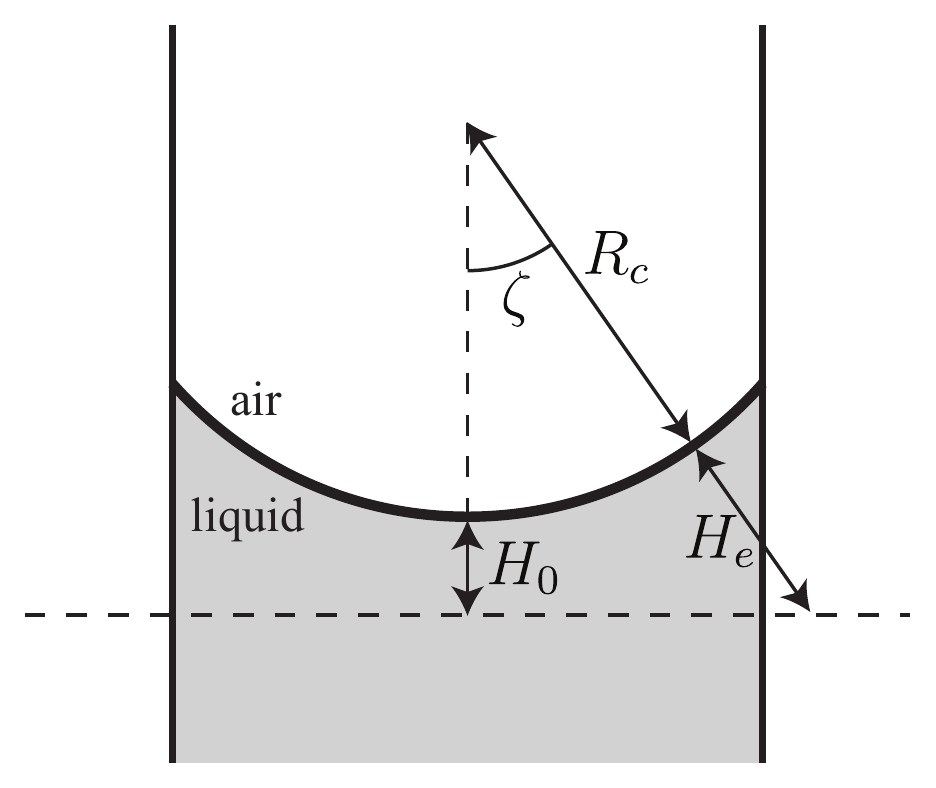}}
  \caption{Definition of the distance $H_0$, the effective distance $H_e$, radius of curvature $R_c$ and angle $\zeta$.}\label{fig:model}
\end{figure}

The free surface is a curved equipotential surface that needs to matched to the plane equipotential surfaces in the bulk. With reference to figure \ref{fig:model} we introduce a distance $H_0$, ultimately to be treated as a fitting parameter, as the smallest distance from the free surface where we assume the equipotential surfaces to be unaffected by the curved interface. We now calculate the distance between the free surface and the horizontal plane defined by $H_0$ in the direction normal to the free surface. We will call this the $\zeta$-dependent effective distance $H_e$:
\begin{equation}
    H_e(\zeta) = \frac{R_c+H_0}{\cos\zeta}-R_c
\end{equation}
where we have defined the position on the free surface as a function of the radius of curvature $R_c$ and the angle $\zeta$, as shown in figure~\ref{fig:model}.
Because the potential difference between the plane defined by $H_0$ and the free surface is constant, we expect the velocity to be inversely proportional to the effective distance $H_e$:
\begin{equation}
    V_{fs}(\zeta)=\frac{A}{H_e(\zeta)}
    \label{eq:meniscus_velocity}
\end{equation}
with $A$ a constant that we will determine using volume conservation: The flux through a cross-section in the bulk, where the velocity is uniform, must equal the flux through the free surface:
\begin{equation}
    V_B\pi R_t^2=\int_0^{2\pi}\int_0^{\pi/2-\theta}V_{fs}(\zeta)R_c^2\sin\zeta d\zeta d\xi,
    \label{eq:volume_conservation}
\end{equation}
where $V_B=\alpha\frac{\Delta p\Delta t}{\rho\lambda}$, $R_c=R_t/\cos\theta$, and $R_t$ the tube radius. $\alpha$ is a prefactor which should be of order unity, reflecting the one-dimensional character of the flow in the bulk. We expect $\alpha$ to become closer to 1 when $\lambda/R_t$ increases.

We now have an expression for $A$, which we substitute in (\ref{eq:meniscus_velocity}), and we arrive at the following velocity on the free surface $V_0\equiv V_{fs}(\zeta=0,t=\Delta t)$:
\begin{equation}
    V_0 = \alpha \frac{\Delta p\Delta t}{\rho\lambda}
    \frac{1}{2h_0}
    \frac{\cos\theta}{b\log\left(\frac{\sin\theta-b}{1-b}\right) + \sin\theta - 1}
    \label{model_correct_V0}
\end{equation}
with $b=1+h_0\cos\theta$, and the geometrical factor $h_0=H_0/R_t$. The value of $h_0$ only needs to be determined once by fitting, since we do not expect it to change with other parameters.

The maximum velocity remains
\begin{equation}
    V_{max} = V_0(1+\beta\cos\theta),
    \label{model_correct_Vmax}
\end{equation}
with $V_0$ given by (\ref{model_correct_V0}).

\begin{figure}
  \centerline{\includegraphics[width=\textwidth]{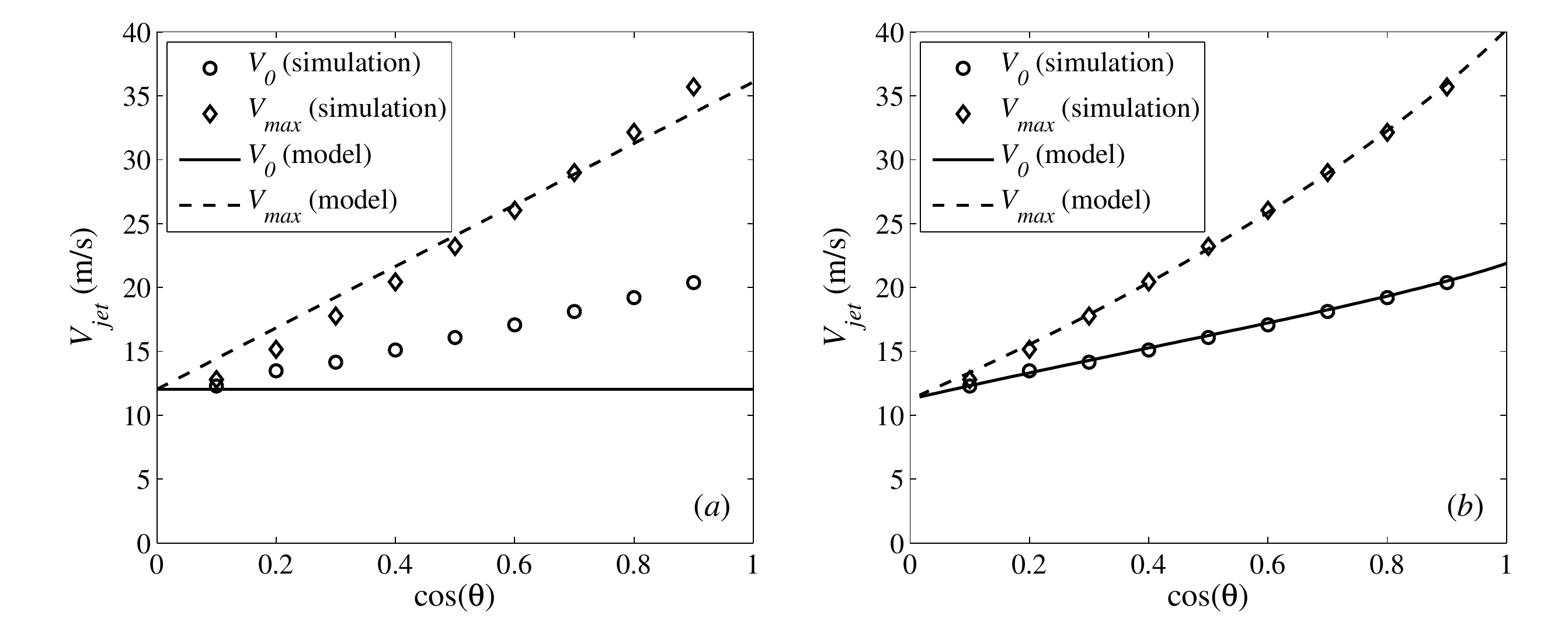}}
  \caption{The velocity $V_0$ of the jet just after the pressure pulse and the maximum velocity $V_{max}$ of the jet. Parameters: $R_{tube}=100~\mathrm{\mu m}$, $\Delta p=3040~\mathrm{bar}$, $\Delta t=50~\mathrm{ns}$, $\lambda=1250~\mathrm{\mu m}$. The diamonds and circles are results from boundary-integral simulations. In (\textit{a}) the solid line corresponds to equation \ref{eq:V_0} and the dashed line corresponds to equation \ref{eq:V_max}, with $\beta=2$. In (\textit{b}) the solid line corresponds to equation \ref{model_correct_V0} and the dashed line corresponds to equation \ref{model_correct_Vmax}, with $\alpha=0.94$, $\beta=0.84$, and $h_0=0.26$.}\label{fig:jetvel_model}
\end{figure}

In figure~\ref{fig:jetvel_model}(\textit{b}) we compare the corrected model to the boundary integral simulations, with $\alpha=0.94$, $\beta=0.88$, and $h_0=0.26$. We find an excellent agreement between the simulations and the model given by (\ref{model_correct_V0}) and (\ref{model_correct_Vmax}). Figure~\ref{fig:theta} shows the comparison of the model with both experiments and simulations, for a different tube radius and bubble distance, but we have used the same value for $h_0$. Note that in the comparison shown in figure~\ref{fig:theta}, $\alpha$ is smaller, which is due to the fact that in that case $\lambda\sim R_t$.

\section{Conclusions and discussion}\label{sec:concl}
We have numerically investigated the formation of microjets in a capillary by laser induced cavitation using axisymmetric boundary integral simulations. Although compressibility plays an important role in the formation and initial growth of the bubble as well as in the subsequent pressure wave that travels through the liquid, we have assumed incompressibility for our numerical simulations. This is possible because the compressibility is only important during the very short period of the pressure wave reflecting from the free surface, which we have modeled by applying a short strong pressure pulse on the bubble inside the capillary. After the initial pressure impulse, the formation of the jet can be considered as incompressible, because the observed speeds are mostly much smaller than the speed of sound in water, and pressures are moderate.

We have found a convincing agreement in shape and evolution of the jet between the simulations and the experiments, which has allowed us to perform a detailed study of the involved parameters, including those which are difficult to access in experiments.

We compared the influence of the different parameters on the maximum achieved velocity of the jet.
We have found good agreement between the simulations and the experiments by investigating the influence of the distance $\lambda$ and contact angle $\theta$. It is however much harder to directly relate the absorbed laser energy in the experiment to the pressure pulse in the simulation. By comparison we were able to derive that the pressure pulse $\Delta p\Delta t$ is linearly related to the absorbed energy $E$, and we have given arguments for the dependence of the pressure on the capillary radius $R_t$.

The jet velocities we find in the simulations can be reproduced accurately by a simple model. We developed this model starting with a one-dimensional approximation for the pressure pulse and dimensional analysis for the focusing effect during jet formation. We improved the one-dimensional approximation by making a correction for the curved interface during the short pressure pulse, where deformation of the meniscus can be neglected.

\begin{acknowledgments}
We would like to thank Claas Willem Visser, Sascha Hilgenfeldt and Leen van Wijngaarden for helpful discussions.
I.P. especially thanks Stephan Gekle for his assistance with the BI code, and acknowledges NWO for financial support.
\end{acknowledgments}

\bibliographystyle{jfm}

\bibliography{microjet}

\end{document}